\documentclass[twocolumn,showpacs,amsmath,amssymb,pra]{revtex4}
\usepackage[latin9]{inputenc}
\usepackage{textcomp}
\usepackage{amsmath}
\usepackage{amssymb}
\usepackage{graphicx}
\usepackage{epstopdf}
\usepackage{bbm}
\usepackage[toc,page,title,titletoc,header]{appendix}
\usepackage{dcolumn}\usepackage{bm}
\usepackage{amsfonts}
\usepackage{epsfig}
\usepackage{mathrsfs}
\newcommand{\beq}{\begin{eqnarray}}
\newcommand{\eeq}{\end{eqnarray}}

\newcommand{\be}{\begin{equation}}
\newcommand{\ee}{\end{equation}}
\newcommand{\bey}{\begin{eqnarray}}
\newcommand{\eey}{\end{eqnarray}}
\newcommand{\ba}{\begin{array}}
\newcommand{\ea}{\end{array}}
\newcommand{\bi}{\begin{itemize}}
\newcommand{\ei}{\end{itemize}}
\newcommand{\bem}{\begin{enumerate}}
\newcommand{\eem}{\end{enumerate}}
\newcommand{\bw}{\begin{widetext}}
\newcommand{\ew}{\end{widetext}}
\newcommand{\ra}{\rangle}
\newcommand{\la}{\langle}

\newcommand{\ov}{\overline}

\begin{document}

 \title{Half-width of local spectral density of states given by width of nonperturbative parts
 of eigenfunctions: The Wigner-band-matrix model}

\author{ Yijian Zou  \footnote{Email address: zyj245@mail.ustc.edu.cn}, 
Yuchen Ma \footnote{Email address: mayuchen@mail.ustc.edu.cn},
Peijun Zhu, Jiaozi Wang, and Wenge Wang
\footnote{Email address: wgwang@ustc.edu.cn}
}
\affiliation{
 Department of Modern Physics, University of Science and Technology of China,
 Hefei 230026, China }
 \date{\today}

 \begin{abstract}
   It is shown that, for a Hamiltonian with a band structure, the half width of local spectral
   density of states, or strength function, is closely related to the width of the nonperturbative 
   (NPT) parts of energy eigenfunctions.
   In the Wigner-band random-matrix model, making use of a generalized Brillouin-Wigner perturbation 
   theory, we derive analytical expressions for the width of the NPT parts
   under weak and strong perturbation.
   An iterative algorithm is given, by which the NPT widths can be computed efficiently, and is used
   in numerical test of the analytical predictions.

 \end{abstract}
%05.30.-d Quantum statistical mechanics
%03.65.-w   Quantum mechanics
%03.65.Yz   Decoherence; open systems; quantum statistical methods

 \maketitle

%\tabSPofcontents

 A note in the beginning by the authors:
 In the present version of the draft, the language and expressions have
 not been polished, yet. We are sorry for this and are to revise the draft as soon as possible.

 \section{Introduction}

 The so-called local spectral density of states (LDOS), also known as strength function in nuclear
 physics, is given by projection of an unperturbed state in perturbed
 states as a function of the energy difference between the unperturbed and the perturbed energies.
 This quantity is useful in the study of many properties of systems, for example,
 in the study of relaxation processes, as well as of various transition probabilities and echoes.
 Among properties of the LDOS that are of relevance, of particular interest is
 its half-width \cite{Lau95,Jac97,Carlos98,Flm,Paski12,Santos12,Santos14}.
 However, except in the case of weak perturbation, analytical study of the width of LDOS is
 usually quite difficult and not much is known about its generic properties.

 As well known, random matrices are useful in the study of complex quantum systems.
 For example, the relation has been established between
 statistical properties of the spectra of quantum chaotic systems and those of full random matrices
 such as Gaussian orthogonal ensembles (GOE).
 A big difference between such random matrices and realistic systems lies in diagonal
 elements of the matrixes.
 For this reason, Wigner proposed to consider the so-called Wigner-band random-matrix (WBRM) model,
 in which the Hamiltonian matrices have increasing diagonal elements and random off-diagonal
 elements within a band \cite{WBRM}.
 This model is regarded as being of relevance in the study of atomic nuclei,
 cold atoms, and disordered systems.
 Analytical study of the WBRM model is much more difficult than that for full random matrices
 such as GOE.

 A useful method of studying properties of energy eigenfunctions (EFs)
 under non-weak perturbation is given by
 a generalization of the Brillouin-Wigner perturbation theory (GBWPT) \cite{pre-98},
 particularly, for Hamiltonian matrices with band structure.
 The GBWPT shows that an EF can be divided into a non-perturbative(NPT) part
 and a perturbative(PT) part, the latter of which can be expanded in a convergent
 perturbation expansion by making use of the former.
 For a Hamiltonian matrix with a band structure, loosely-speaking, its EFs have exponential-type
 decay in the PT regions, hence, their main bodies should lie in the NPT regions \cite{pre00}.

 Interestingly, numerical simulations carried out in the WBRM model reveal a relation between
 the width of LDOS and the width of the NPT regions of EFs.
 In this paper, we give further investigation for this phenomenon.
 We first use the GBWPT to explain the numerically-observed relation
 between the width of LDOS and the width NPT regions of EFs.
 Then, we give analytical study of the width of the NPT parts and develop a method
 of studying analytically its variation of with the perturbation strength, from weak to strong.
 By this approach, main behaviors of the width of LDOS can be explained quantitatively.
 We also develop a generic algorithm that can efficiently compute  the
 width of NPT regions of EFs for band matrixes and use this algorithm to test
 our analytical results.

 The paper is structured as follows.
 In Sec.II, we first recall basic results of the GBWPT, giving definition of PT and NPT regions of
 EFs, then, discuss the WBRM model.
 We also explain the relationship between the NPT width and the half width of LDOS in this model.
 In Sec.III, we derive the analytical expressions for the average NPT width in the WBRM
 model for weak and strong perturbations.
 In Sec.IV, we introduce a recursive algorithm to compute the NPT width for band matrices,
 use it to test our analytical results given in Sec.III.

 \section{Theory and Model}
 \label{sect-mainr}

\subsection{Generalized Brillouin-Wigner perturbation theory}
\label{sect-GBWPT}

 In this section, we recall basic contents of the GBWPT.
 Consider a Hamiltonian of the form
\begin{equation}\label{H}
 H(\lambda)=H_{0}+\lambda V,
\end{equation}
 where $H_{0}$ is an unperturbed Hamiltonian and $\lambda V$ represents a perturbation
 with a running parameter $\lambda$.
 Eigenstates of $H(\lambda)$ and $H_{0}$ are denoted by $|\alpha\rangle$ and $|k\rangle$,
 respectively,
\begin{equation}
H(\lambda)|\alpha\rangle=E_{\alpha}(\lambda)|\alpha\rangle,
\quad H_{0}|k\rangle=E_{k}^{0}|k\rangle,
\end{equation}
 with $\alpha$ and $k$ in energy order.
 Components of the EFs are denoted by $C_{\alpha k} = \la k|\alpha\ra$.

 In the GBWPT, for each perturbed
 state $|\alpha\rangle$, the set of the unperturbed states $|k\rangle$
 is divided into two substes, denoted by $S_{\alpha}$ and $\overline{S}_{\alpha}$.
 The related projection operators,
\begin{equation}
P_{S_{\alpha}}=\sum\limits _{|k\rangle\in S_{\alpha}}|k\rangle\langle{k}|,Q_{\overline{S}_{\alpha}}
=\sum\limits _{|k\rangle\in\overline{S}_{\alpha}}|k\rangle\langle k|=1-P_{S_{\alpha}},
\end{equation}
 divide the perturbed state into two parts,
$|\alpha_{s}\rangle\equiv{P_{S_{\alpha}}|\alpha\rangle}$,
$|\alpha_{\overline{s}}\rangle\equiv Q_{\overline{S}_{\alpha}}|\alpha\rangle$.
 As shown in Ref.\cite{pre-98}, if the above-discussed division satisfies the following condition, namely,
\begin{equation}
 \lim _{n \to \infty }  \langle \phi |(T_{\alpha }^{\dagger})^n
T_{\alpha }^n |\phi  \rangle =0 \quad \forall \phi,  \label{conv}
\end{equation}
 where
\begin{equation}\label{T-alpha}
T_{\alpha}=\frac{1}{E_{\alpha}-H_{0}}Q_{\overline{S}_{\alpha}}\lambda{V},
\end{equation}
 then, making use of the part $|\alpha_{s}\rangle$,
the other part $|\alpha_{\overline{s}}\rangle$ can be expanded
in a convergent perturbation expansion, i.e.,
\begin{equation}\label{alpha-ovs}
|\alpha_{\overline{s}}\rangle=T_{\alpha}|\alpha_{s}\rangle+T_{\alpha}^{2}|\alpha_{s}\rangle
+\cdots+T_{\alpha}^{n}|\alpha_{s}\rangle+\cdots.
\end{equation}

 Let us consider an operator $W_{\alpha }$ in the subspace spanned by unperturbed states
 $|k\ra \in {\overline S}_{\alpha } $, namely,
\be W_{\alpha } := Q_{{\overline S}_{\alpha }} V \frac{1}{E_{\alpha }-H^0} Q_{{\overline S}_{\alpha }},
\label{U} \ee
 and use $|\nu\rangle $ and $w_{\nu }$ to denote its eigenvectors
 and eigenvalues, $W_{\alpha } |\nu\rangle = w_{\nu } |\nu\rangle $, where for brevity we omit the subscript
 $\alpha$ for $|\nu\rangle $ and $w_{\nu }$.
 It is easy to verify that the condition (\ref{conv})  is equivalent to the requirement that
\begin{equation} \label{inequ}
 |\lambda w_{\nu }| < 1 \quad \forall |\nu\rangle.
\end{equation}

 In a quantum chaotic system $H(\lambda)$,
 all good quantum numbers of the unperturbed system $H_0$ have been destroyed,
 except that related to the energy.
 Therefore, in the study of statistical properties of the EFs,
 we consider those sets $S_\alpha$, each corresponding to a connected region in the unperturbed energy, namely,
\begin{equation}\label{S-alpha}
 S_\alpha = \{ |k\ra : k_1 \le k \le k_2 \}.
\end{equation}
 Among the sets $S_\alpha$ for which Eq.(\ref{conv}) is satisfied, the most important is the smallest one.
 We call the smallest set $S_{\alpha}$, under the condition (\ref{conv}),
 the \emph{non-perturbative (NPT) region} of the state $|\alpha\ra$ and, correspondingly,
 the set $S_{\ov \alpha}$ the \emph{perturbative (PT) region}.
 Clearly, the NPT region of $|\alpha\ra$ has the smallest value of $(k_2-k_1)$.
 Below, we use $N_p$ to denote the width of the NPT region, namely,
\begin{equation}\label{}
 N_p = k_2-k_1.
\end{equation}

 In the case that $\lambda$ is sufficiently small and $E_\alpha$ is not close to
 the unperturbed eigenenergies,
 the condition (\ref{conv}) can be satisfied with a set $S_{\alpha}$ including
 only one unperturbed state $|k_0\ra$, whose energy $E^0_{k_0}$ is the closest to $E_\alpha$.
 In this case, $k_1=k_2 = k_0$.
 With increasing perturbation strength $\lambda$, usually the width the NPT region increases.

 As an application of the GBWPT, we discuss a Hamiltonian that has a matrix with
 a band structure in the unperturbed basis.
 Let us expand the state vector $Q_{{\overline S}_{\alpha }}\lambda V |\alpha _s\rangle $
 in the basis $|\nu\rangle $, giving
\begin{equation}\label{Qas-expan}
 Q_{{\overline S}_{\alpha }}\lambda V |\alpha _s\rangle = \sum_{\nu } d_{\nu } |\nu\rangle .
\end{equation}
 Substituting Eq.(\ref{T-alpha}) and Eq.(\ref{Qas-expan}) into Eq.(\ref{alpha-ovs}),
 after simple derivation, it is found that, for each unperturbed state $|j\rangle $
 in the set ${\overline S}_{\alpha }$,
 the component $C_{\alpha j}=\langle j|\alpha \rangle $ is written as
\begin{equation}
 C_{\alpha j} = \frac 1{E_{\alpha }-E^0_j} \sum_{\nu }
\left [ \frac {d_{\nu }}{1- \lambda w_{\nu}} \langle j|\nu\rangle \right ]
\left ( \lambda w_{\nu } \right )^{m-1} , \label{CU}
\end{equation}
 where $m$ is the smallest positive integer for
 $\langle j |(Q_{{\overline S}_{\alpha }}V)^m |\alpha _s\rangle $ not equal to zero,
 i.e., the smallest steps for $|j\ra$ to be coupled to $|\alpha_s\ra$ through $V$ \cite{wwg-GBWPT}.
 Consider a Hamiltonian matrix with a band structure discussed above,
 specifically, with a band width $b$, that is, $\la k| V|k'\ra =0$ if $|k-k'| >b$.
 Let us consider $C_{\alpha j}$ of $j>k_2$.
 It is easy to see that $m \ge \frac{1}{b}(j-k_2)$.
 Since $|\lambda w_{\nu }| < 1$, Eq.(\ref{CU}) shows that the EF has an exponential-type decay
 with increasing $j$.
 Similarly, the EF has an exponential-type decay with decreasing $j$ for $j<k_1$.

 It is seen that $m=1$ for $j$ in the two regions $[k_1-b,k_1]$ and $[k_2,k_2+b]$.
 According to Eq.(\ref{CU}), the exponential-type decay does not appear in these two regions.
 We call them the \emph{shoulders} of the NPT region.
 Clearly, the main body of the EF should lie within the region $[k_1-b,k_2+b]$, namely,
 in the NPT-plus-shoulder region.

\subsection{The WBRM model}

 In the WBRM model, one considers a perturbed Hamiltonian matrix written in the form in Eq.(\ref{H}).
 Here, the unperturbed Hamiltonian $H_0$ takes a diagonal form with $E^0_i=i$  ($i=1 \cdots ,N$).
 The elements $v_{ij}$ of the perturbation $V$ are random numbers with Gaussian distribution for
 $1\leqslant |i-j| \leqslant b$ $(\langle v_{ij} \rangle =0, \langle v^2_{ij} \rangle=1$)
 and are zero otherwise.
 Thus, the Hamiltonian matrix has a band structure with a bandwidth $b$.
 At large $\lambda$, the EFs have the feature of localization in the energy shell \cite{CCGI96}.

 It proves convenient to introduce a matrix
\begin{equation}\label{Udef}
  U=Q\frac{1}{E_\alpha-H_0}\lambda VQ,
\end{equation}
 where $Q$ is a projection operator introduced in the previous section, with
 the subscript $\alpha_s$ omitted.
 Elements of $U$ are
\begin{equation}\label{Uij}
  U_{ij}=\left\{
\begin{array}{cc}
0,&p_1\le i,j \le p_2,\\
\displaystyle{\frac{\lambda V_{ij}}{E_\alpha-E^0_i}},& \rm{otherwise.}
\end{array}
\right.
\end{equation}
 We use $s(A)$ to denote the the maximum of the modulus of the eigenvalues of an operator $A$.
 For example, $s(U)$ is the maximum of $|u_m|$, where $u_m$ are eigenvalues of $U$.
 Then, the condition (\ref{conv}) is equivalent to the following requirement, namely,
\begin{equation}\label{}
  s(U) <1.
\end{equation}

\subsection{Half width of LDOS and NPT width}

\begin{figure}
\includegraphics[width=0.48\textwidth]{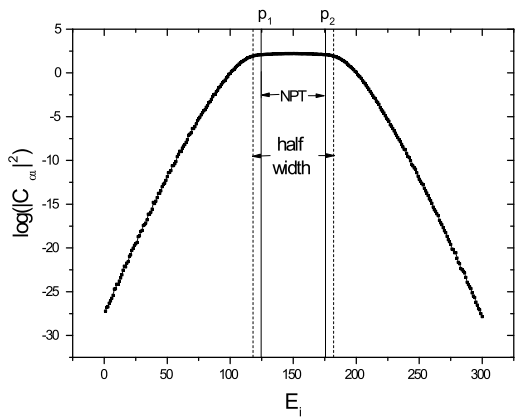}
% Here is how to import EPS art
\caption{Average shape of eigenfunctions of Wigner-band random matrices, $b=8$,$\lambda=5$,$N=500$.
The shape is steady inside NPT region but decreases exponentially outside shoulders.
Half width of eigenfunctions is near NPT width plus two shoulders.}
\label{avereig}
\end{figure}

\begin{figure}
\includegraphics[width=0.48\textwidth]{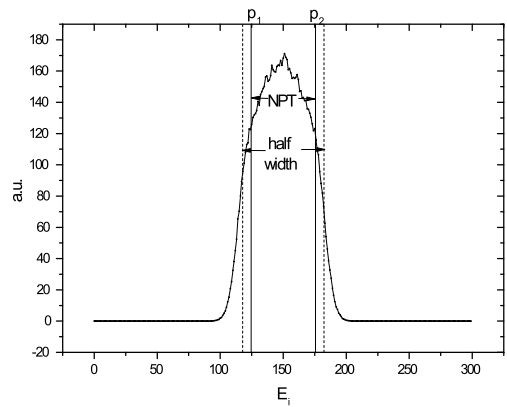}
% Here is how to import EPS art
\caption{Average shape of eigenfunctions of Wigner-band random matrices, $b=8$,$\lambda=5$,$N=500$.
}
\label{avereig2}
\end{figure}

 The shape of an EF of $|\alpha\ra$ can be written in the form
\begin{equation}\label{EF}
 \rho_{F} = \sum_k|C_{\alpha k}|^2\delta(E_k^0 - E_\alpha).
\end{equation}
 Its averaged shape of EFs, denoted by $\Pi$,
 can be obtained by taking average of $\rho_{F}$ over different EFs,
 with the energies $E_\alpha$ shifted to a common position, say, to the origin of $E=0$.
 We use $\varepsilon_{\alpha k}$ to denote the energy difference $(E_\alpha - E_k^0)$,
 namely, $\varepsilon_{\alpha k} = E_\alpha - E_k^0$.
 The averaged EF is a function of $\varepsilon$ and is written as $\Pi(\varepsilon)$.
 In the WBRM, whose Hamiltonian has a band width b, as discussed previously, main bodies of the EFs
 lie within the NPT+shoulder regions.
 Therefore, the width of the averaged EF, denoted by $w_F$, satisfies
\begin{gather}\label{}
 w_F \lesssim N_p +2b.
\end{gather}

 Similarly, the LDOS of unperturbed state $|k\ra$ is written as
\begin{equation}\label{EF}
 \rho_{LDOS}(E_k^0,E_\alpha) = \sum_\alpha |C_{\alpha k}|^2 \delta(E_\alpha - E_k^0).
\end{equation}
 The averaged LDOS, obtained with $E_k^0$ moved to $E^0=0$, is a function of $\varepsilon$
 and is indicated by $\rho_L(\varepsilon)$.

 We use $w_L$ to denote the half-width of the averaged LDOS $\rho_L(\varepsilon)$.
 Numerically, we found that in the WBRM model the averaged EFs almost fully fill the NPT region
 (see Figs.\ref{avereig} and \ref{avereig2}).
 Hence, we have
\begin{gather}\label{wL-Np}
 w_L \approx N_p + 2\eta b,
\end{gather}
 where $\eta \sim 1$ determined by decay of the EFs outside their NPT regions.
 Thus, $w_L$ can be estimated, once the NPT-region width $N_p$ is known.

\section{Width of NPT regions for weak and strong perturbation}\label{sect-analytical}

 In this section, we discuss variation of the NPT-region $N_p$ with the perturbation strength
 $\lambda$.

\subsection{NPT width for $b=1$ at small $\lambda$}

 At $b=1$, elements of the matrix $U$ in Eq.(\ref{Uij}) have the simple expression,
 $U_{ij}={\lambda V_{ij}}\delta_{i,j\pm 1}/{(E_\alpha-E^0_i)}$ for $i,j$ outside of
 the interval $[p_1,p_2]$.

 When $\lambda$ is quite small, one usually has $p_1=[E_\alpha]$ and $p_2=[E_\alpha]+1$,
 which gives $N_p=1$.
 In some special realization of the random numbers for off-diagonal elements of the Hamiltonian,
 one may have $N_p=2$.
 To compute $s(U)$, let us consider five basis states $|k\ra$ with unperturbed energies $E^0_k$
 just above $E_\alpha$, as well as five basis states with $E^0_k$ just below $E_\alpha$.
 Truncate the matrix $U$ in these basis states, one gets the following two five-dimensional
 sub-matrices, denoted by $U_{\rm{up}}$ and $U_{\rm{down}}$, namely,
\begin{widetext}
\begin{equation}
  U_{\rm{up}}=\lambda\left(\begin{array}{ccccc}0&\displaystyle\frac{V_{p_1-4,p_1-5}}{E_\alpha-p_1+5}&0 &0 &0 \\
  \displaystyle\frac{V_{p_1-4,p_1-5}}{E_\alpha-p_1+4}&0&\displaystyle\frac{V_{p_1-3,p_1-4}}{E_\alpha-p_1+4}&0 &0 \\
  0&\displaystyle\frac{V_{p_1-3,p_1-4}}{E_\alpha-p_1+3}&0&\displaystyle\frac{V_{p_1-2,p_1-3}}{E_\alpha-p_1+3}& 0 \\
    0 &0 &\displaystyle\frac{V_{p_1-2,p_1-3}}{E_\alpha-p_1+2}&0&\displaystyle\frac{V_{p_1-1,p_1-2}}{E_\alpha-p_1+2}\\
     0 &0 & 0&\displaystyle\frac{V_{p_1-1,p_1-2}}{E_\alpha-p_1+1}&0 \end{array}\right),
\end{equation}
\begin{equation}
  U_{\rm{down}}=-\lambda\left(\begin{array}{ccccc}0&\displaystyle\frac{V_{p_2+2,p_2+1}}{p_2-E_\alpha+1}&0 &0 &0 \\
   \displaystyle\frac{V_{p_2+2,p_2+1}}{p_2-E_\alpha+2}&0&\displaystyle\frac{V_{p_2+3,p_2+2}}{p_2-E_\alpha+2}&0 &0 \\
    0&\displaystyle\frac{V_{p_2+3,p_2+2}}{p_2-E_\alpha+3}&0&\displaystyle\frac{V_{p_2+4,p_2+3}}{p_2-E_\alpha+3}&0  \\
    0 & 0&\displaystyle\frac{V_{p_2+4,p_2+3}}{p_2-E_\alpha+4}&0&\displaystyle\frac{V_{p_2+5,p_2+4}}{p_2-E_\alpha+4}\\
    0  & 0& 0&\displaystyle\frac{V_{p_2+5,p_2+4}}{p_2-E_\alpha+5}&0 \end{array}\right).
\end{equation}
\end{widetext}
 Since elements of $U$ outside the above two matrices are
 generally much smaller than those inside them,
 $s(U)$ can be approximated by the maximal modulus of the eigenvalues of the
 two sub-matrices, i.e., $s(U)=max\{s(U_{\rm{up}}),s(U_{\rm{down}})\}$.
 Since the two matrices $U_{\rm{up}}$ and $U_{\rm{down}}$ have similar structures,
 we can focus on $s(U_{up})$ only.

 Below, we give a condition under which $s(U_{up})>1$.
 For the sake of convenience in presentation, here we introduce some notations
 that will be frequently used,
\begin{gather}\label{}
 f=E_\alpha-p_1+1, \quad g=p_2-E_\alpha+1.
\end{gather}
 Using $r_\alpha$ to indicate the decimal part of $E_\alpha$,
 taking $N_p=1$, one has $f=1+r_\alpha,g=2-r_\alpha$.
 We use $v_i$ of $i=1,2,3,4$ to denote the four nonzero elements $V_{p,p-1}$ in
 the matrix $U_{\rm{up}}$ from top to bottom.

 Direct computation shows that the eigenvalues $\mu$ of $U_{\rm{up}}$ satisfy the following equation,
\begin{equation}\label{quad}
  h(\mu)=\mu^4-A\mu^2+B=0,
\end{equation}
 where
\begin{eqnarray}\label{AA}
  A&=&\lambda^2[\frac{v^2_1}{(f+3)(f+4)}+\frac{v^2_2}{(f+2)(f+3)} \nonumber \\
  &+&\frac{v^2_3}{(f+1)(f+2)}+\frac{v^2_4}{f(f+1)}],
\end{eqnarray}
\begin{eqnarray}\label{BB}
  B&=&\lambda^4[\frac{v^2_1 v^2_3}{(f+1)(f+2)(f+3)(f+4)} \nonumber \\
  &+&\frac{v^2_2 v^2_4}{f(f+1)(f+2)(f+3)} \nonumber \\
  &+&\frac{v^2_1 v^2_4}{f(f+1)(f+3)(f+4)}].
\end{eqnarray}
 It is easy to verify that $A^2-4B>0$.
 Thus, we get the solution
\begin{equation}
  \mu^2_\pm=\frac{A\pm\sqrt{A^2-4B}}{2}.
\end{equation}
 Therefore, $s(U_{\rm{up}})>1$ is equivalent to $\mu^2_+>1$.
 Further computation shows that $s(U_{\rm{up}})>1$ is equivalent to the condition that
 $A>2$ or $A>B+1$.

% Denote the four additive terms in medium bracket in Eq.(\ref{AA}) as $a_1,a_2,a_3,a_4$.
% We may directly examine the discriminant of Eq.(\ref{quad}) that
%\begin{eqnarray}
%  A^2-4B&>&A^2-4B-4a_2a_3 \nonumber \\
%  &=&(a_2-a_3)^2+(a_1-a_4)^2-2(a_2-a_3)(a_1-a_4) \nonumber \\
%  &>&0.
%\end{eqnarray}

 Detailed analysis shows that, for $\lambda \lesssim 1$, usually one has $A<2$ and $B\ll1$.
 In fact, as an estimation, taking $f \approx 1$ and $v_i \approx 1$,
 one has $A \approx 4\lambda^2/5$ and $B \approx 3\lambda^4/40$.
 Then, the condition $s(U_{up})>1$ is simplified to $A>1$, i.e.,
\begin{eqnarray}
   X_{up} > \frac{1}{\lambda^2},
\end{eqnarray}
 where
\begin{eqnarray}
   X_{up}=\frac{v^2_1}{(f+3)(f+4)}+\frac{v^2_2}{(f+2)(f+3)} \nonumber \\
   +\frac{v^2_3}{(f+1)(f+2)}+\frac{v^2_4}{f(f+1)}.
\end{eqnarray}
 Similarly, we have the condition $X_{\rm{down}}>1/\lambda^2$ for $s(U_{\rm{down}})>1$,
 where $X_{\rm{down}}$ has the same expression as $X_{up}$ with $f$ replaced by $g=3-f$.
 Denoting $X=max\{X_{\rm{up}},X_{\rm{down}}\}$,
 it is seen that the condition $s(U)>1$ is equivalent to that $X>1/\lambda^2$.

\begin{figure}
\includegraphics[width=0.48\textwidth]{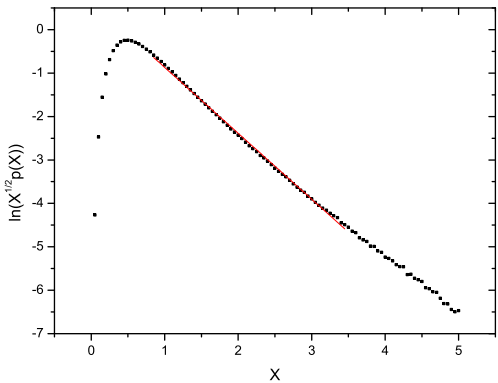}
% Here is how to import EPS art
\caption{Fitting the distribution function of $X$ with $p(X)=CX^{-1/2}e^{-\beta X}(X\ge 1)$,
where $X>1/\lambda^2$ implies NPT width is larger than $1$.}
\label{Xdistr}
\end{figure}

 We have performed a Monte Carlo simulation for the distribution of $X$ and
 found that it can be fitted well by $p(X)=CX^{-1/2}e^{-\beta X}(X\ge 1)$ in the region of interest.
 In fact, the probability for large $X$ is low.
 Our fitting result is shown in Fig.\ref{Xdistr}.

 Now we calculate the average width of the NPT region.
 The probability for $N_p=1$ is given by
\begin{equation}
  P_1=1-\int_{1/\lambda^2}^{+\infty}p(X)dX.
\end{equation}
 Neglecting the small possibility for $N_p >2$ at small $\lambda$,
 the probability of $N_p =2$, denoted by $P_2$, is given by $P_2 = 1-P_1$.
 Then, we have
\begin{equation}
  \la N_p \ra=1+\int_{1/\lambda^2}^{+\infty}p(X)dX.
\end{equation}
Completing the integration, for $\lambda\lesssim 1$, we obtain
\begin{equation}\label{small}
   \la N_p \ra=1+C\sqrt{\frac{\pi}{\beta}}\rm{erfc}(\sqrt\frac{\beta}{\lambda^2}),
\end{equation}
where
\begin{equation}
  \textmd{erfc}(x)=1-\frac{2}{\sqrt{\pi}}\int_0^x e^{-t^2}dt
\end{equation}
is the complementary error function.

\subsection{NPT width for $b=1$ at large $\lambda$}

 Next, we move to the case of large $\lambda$, in which $N_p$ is large.
 Still, for a given value of $\lambda$, due to the random nature of the off-diagonal elements of
 the Hamiltonian, $N_p$ has different values under different realizations of the
 off-diagonal elements.
 Thus, we need to study the probabilities for $N_p$ to take different values.

 As discussed above, the value of $N_p$ is determined by properties of $s(U)$,
 and the matrix $U$ is split into an upper part $U_{\rm{up}}$
 and a lower part $U_{\rm{down}}$.
 Thus, $s(U)=max\{s(U_{\rm{up}}),s(U_{\rm{down}})\}$.
 For $N$ large and $E_\alpha$ in the middle energy region, since the two sub-matrices
 have similar structure, we can consider one of them only, say, $s(U_{\rm{up}})$.
 For brevity, we use $s(U)$ to indicate this sub-matrix.
 For large $\lambda$, one has $p_2-E_\alpha \simeq E_\alpha-p_1 \simeq N_p/2$.

 Let us study the probability of $s(U)<1$ for a given value of $N_p$.
 For later convenience, we write this probability as $1-P(N_p)$.
 Because the NPT width corresponds to the minimum value of $N_p$,
 for which $s(U)<1$, the probability that NPT width is $n$ is $(1-P(n))-(1-P(n-1))=P(n-1)-P(n)$.
 Then,
 \begin{eqnarray}\label{Np}
  \la N_p \ra&=&\sum_{n=1}^{\infty}n(P(n-1)-P(n)) \nonumber \\
  &=&\sum_{n=1}^{\infty}P(n)-\lim_{n\rightarrow\infty}nP(n).
\end{eqnarray}

\begin{figure}
\includegraphics[width=0.48\textwidth]{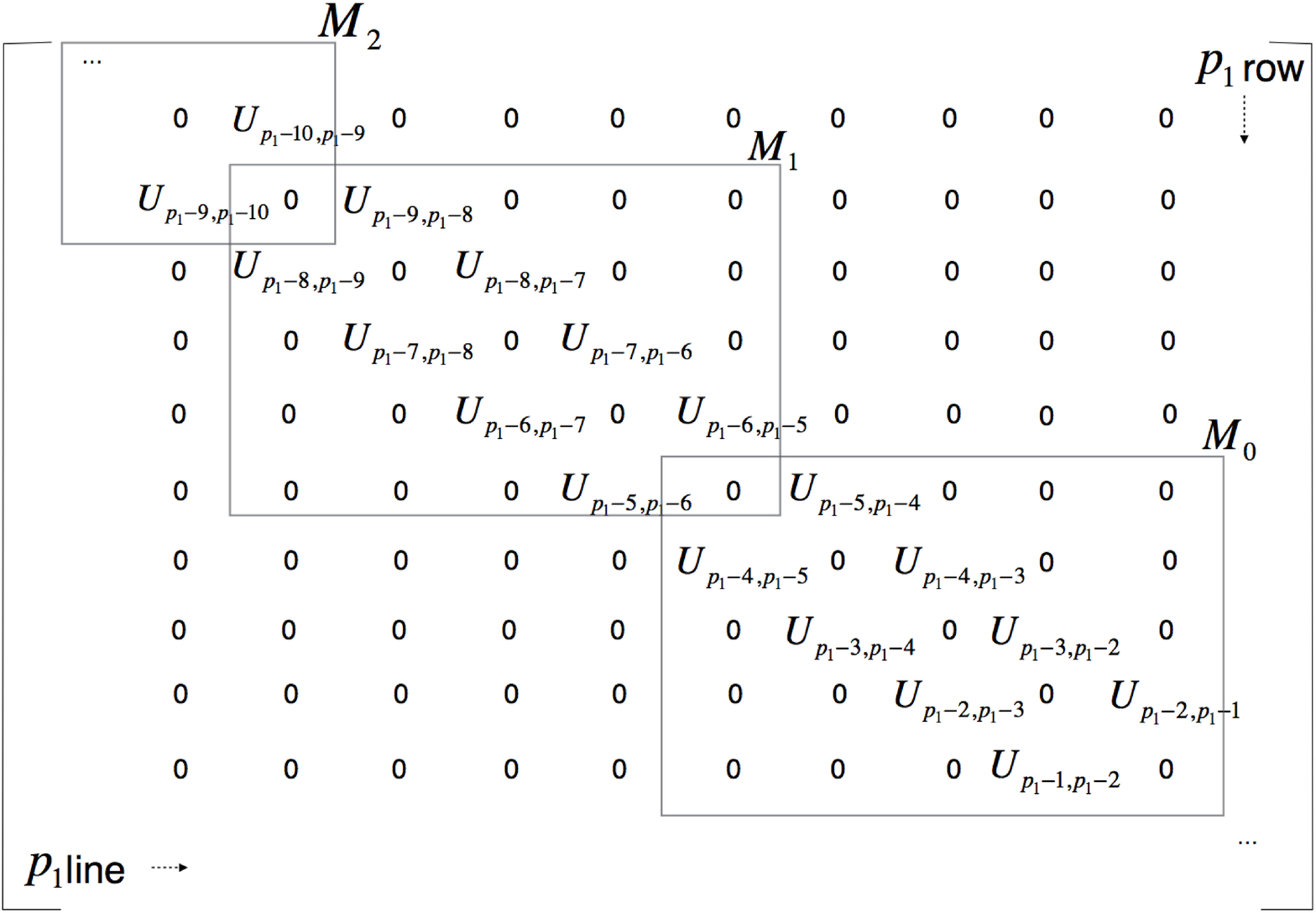}
% Here is how to import EPS art
\caption{A series of matrices truncated from $U$ to estimate the maximum modulus of
eigenvalues of $U$.}
\label{trunc}
\end{figure}

 Next we derive an approximate expression for $P(n)$.
 For this purpose, let us first estimate $s(U)$.
 Consider a series of $m-$dimensional sub-matrices truncated from $U$,
 denoted by $M_i,i=0,1,2,\cdots,n$.
 An illustration of our truncation method is given in Fig.\ref{trunc}.
 In the case of $b=1$, we take $m=5$.
 Generally, we require that $b<m\ll \lambda<N$.

\begin{figure}
\includegraphics[width=0.48\textwidth]{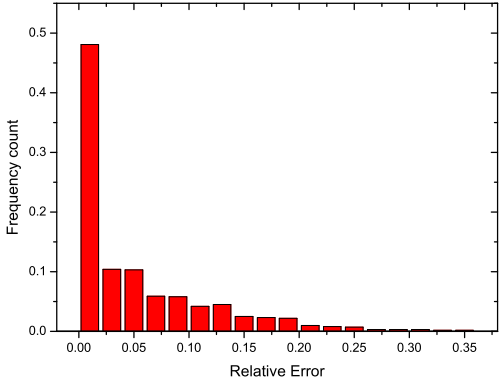}
% Here is how to import EPS art
\caption{Error distribution of estimation in Eq.(\ref{rhoU})}
\label{errorfig}
\end{figure}

 Specifically, sub-matrix $M_i$ is obtained by truncating $U$ from its $l_i$-th row to
 its $l_{i+1}$-th row, and from its $l_i$-th column to its $l_{i+1}$-th column,
 where $l_i=p_1-(m-1)i-1$.
 Elements of $M_i$ are given by
\begin{equation}
  M_{i,pq}=U_{l_{i+1}+p-1,l_{i+1}+q-1}.
\end{equation}
 Thus, the sub-matrices ${M_i}$ contain all nonvanishing elements of $U$, and they
 are independent of each other.
 We found that $|s(U) - \max_{0 \le i \le n}s(M_i)|$ are not large (see  Fig.\ref{errorfig}),
 hence, use the following approximation,
\begin{equation}\label{rhoU}
  s(U) \approx \max_{0 \le i \le n}s(M_i).
\end{equation}

 In the nonzero elements of $U$, the factors $1/(E_\alpha-E^0_i)$ can be regarded as
 a constant for each $M_i$.
 Indeed, in the case of $i<p_1$ and $|i-j|<m\ll  N_p $, one has
\begin{equation}\label{approximation}
  \left|\frac{1}{E_\alpha-E^0_i}-\frac{1}{E_\alpha-E_{j}}\right|
  \approx\frac{m}{(E_\alpha-E^0_i)^2}\approx \frac{m}{N^2_p},
\end{equation}
 Introducing $a_i=E_\alpha-E_{p_1}+(m-1)i+1$ and noting Eq.(\ref{approximation}),
 it is seen that the matrices $M$=$a_i M_i/\lambda$ can be regarded as realizations 
 of the same independent random variables, independent of the label $i$.
 We call $M$ the standard $m$-dimensional matrix.
 It is easy to verify that $M$ is approximately a hermitian matrix and that
 its elements in upper triangle are
\begin{equation}
  M_{ij} \approx  v_i\delta_{i,j-1} \ (i\leq j),
\end{equation}
 where $v_i$ are independent, normally-distributed random variables with mean $0$ 
 and variance $1$.

 We use $h(x)$ to denote the distribution function of $s(M)$,
 and $H(x)$ for the corresponding cumulative distribution function, with
{$H(x) = \int_0^x h(x')dx'$}.
 For brevity, we use $s_M$ to denote $s(M)$.
According to definition of $M$, $M_i=\lambda M/a_i$, then $s(M_i)=\lambda s_M/a_i$.
Then making use of Eq.(\ref{rhoU}), we obtain
\begin{equation}\label{rhoUspecific}
  s(U)=\max_{0 \le i \le n}(\lambda s_M/a_i).
\end{equation}
 We denote the distribution function of $s(U)$ by $w(s)$ and the
corresponding cumulative distribution function by $W(s)$. Using Eq.(\ref{rhoUspecific}), we obtain
\begin{equation}\label{Wrho}
  W(s)=\prod_{i=0}^{n}H(\frac{a_i s}{\lambda}).
\end{equation}
 %As $m\ll \lambda$, we can regard $a_i/\lambda$ as a quasi-continuous variable,
 Let us first take logarithm for both sides of Eq.(\ref{Wrho}),
then, approximate the summation on the right hand side by integration and obtain
\begin{equation}\label{lnWrho}
  \ln W(s)\approx\frac{\lambda}{m-1}\int_{a_0/\lambda}^{+\infty}\ln H(ts)dt,
\end{equation}
where the upper limit is set $+\infty$ because we can regard the matrix as sufficient large.
Note that $a_0\approx N_p/2$, thus we obtain from Eq.(\ref{lnWrho}) that
the probability of $s(U)<1$ is given by
\begin{equation}\label{Prho}
  \exp\left(\frac{\lambda}{m-1}\int_{N_p/2\lambda}^{+\infty}\ln H(t)dt\right).
\end{equation}
 Then,
\begin{equation}\label{Pn}
  P(n)=1-\exp\left(\frac{\lambda}{m-1}\int_{n/2\lambda}^{+\infty}\ln H(t)dt\right).
\end{equation}
 For $n\ll \lambda$, we have $P(n)\approx 1$.
 For $n$ increasing beyond $\lambda$, $P(n)$ decreases and approaches $0$.

\begin{figure}
\includegraphics[width=0.48\textwidth]{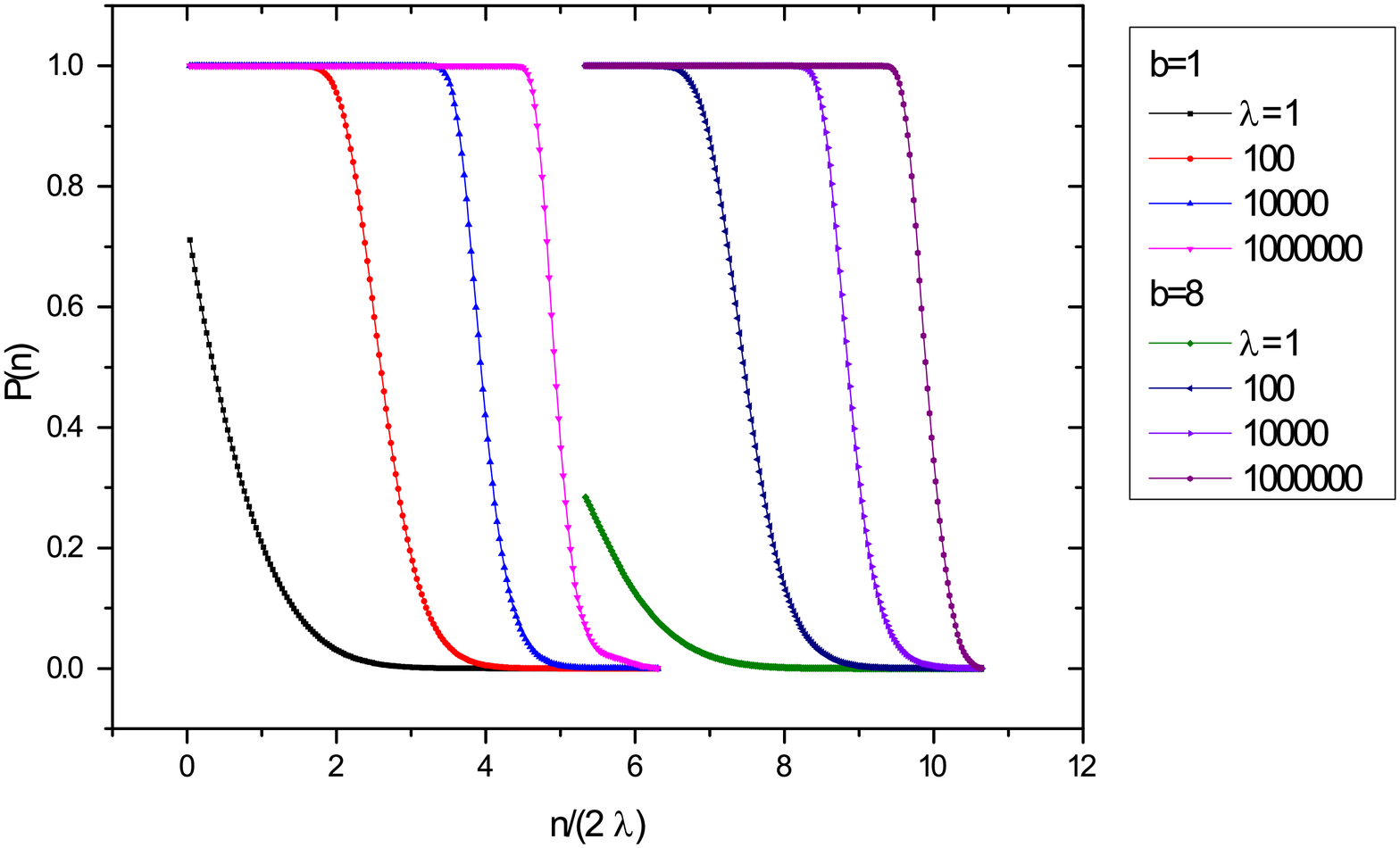}
% Here is how to import EPS art
\caption{Curves of $P(n)$ as $b$ and $\lambda$ vary. We see that when $\lambda$ us sufficiently large,
the curve is like heaviside for both $b$.}
\label{Pnvalue}
\end{figure}

 To evaluate the right hand of Eq.(\ref{Pn}), we need to make use of our
 numerical results of $h(t)$ (see Appendix A).
%  and method of complex trapezoidal integration.
 As shown in Fig.\ref{Pnvalue}, $P(n)$, as a function of $n/2\lambda$,
 has a "ladder" shape when $\lambda$ is very large, and
 $nP(n)\rightarrow0$ as $n\rightarrow\infty$.
 Then, from Eq.(\ref{Np}) and Eq.(\ref{Pn}), we have
\begin{equation}\label{npnc}
  \la N_p \ra=n_c,
\end{equation}
 where $n_c$ is the point at which $P(n_c)=1/2$.
Using Eq.(\ref{Pn}), we obtain
\begin{equation}\label{nc}
 \int_{n_c/2\lambda}^{+\infty}\ln H(t)dt=-\frac{(m-1)\ln 2}{\lambda}.
\end{equation}
As $\lambda$ is large, the absolute value of the right side of Eq.(\ref{nc}) is small,
 hence, $n_c$ should be large. Discussions given in Appendix A show that
\begin{equation}\label{integral}
  \int_x^{+\infty} \ln H(t)dt\sim -\frac{e^{-ax^2}}{x^2}(x\rightarrow +\infty).
\end{equation}
 Then, we get
\begin{equation}\label{large}
  \la N_p \ra \simeq C'\lambda\sqrt{\ln\lambda}(\lambda\gg 1).
\end{equation}
 When $\lambda$ is not extremely large, one has $\la N_p \ra \propto \lambda$.

\subsection{NPT width of Wigner-band random matrix:band width $b\ge 2$}

 In this section, we argue that Eq.(\ref{small}) and Eq.(\ref{large}) are still
 valid for $b\ge 2$, with coefficients in the equations as fitting parameters.

 Let us first discuss the case of small $\lambda$.
 Here, we take the two matrices $U_{\rm{up}}$ and $U_{\rm{down}}$
 [cf.Eqs.(13) and (14)] as the two $5b$-dimensional matrices nearest to $E_\alpha$.
 Neglecting terms of the order of $O(\lambda^2)$ and higher in the eigen-equation,
 we obtain a condition similar to Eq.(\ref{quad}).
 Neglecting $B$ in the eigen-equation, the condition for $N_p=1$ is given by $A>1$.
 Writing $X = A/\lambda^2$, we use $p(X)=CX^{-1/2}e^{-\beta X}$ to
 fit the distribution of $X$.
 Finally, we also get Eq.(\ref{small}), with $C$ and $\beta$ as fitting parameters.

 Next, for large $\lambda$, when deriving Eq.(\ref{large}),
 we need to use properties of $H(t)$ and
 the cumulative distribution function of the maximal eigenvalue of a standard $m$-dimensional
 random matrix. We can choose $m$, such that it is sufficiently larger than $b$ but still
 sufficiently smaller than $\lambda$.
 Then, the asymptotic behavior of $H(t)$ is still given by Eq.(\ref{integral})
 (see Appendix A). Finally, the result Eq.(\ref{large}) remains unchanged,
 with a different parameter $C'$.

\section{Numeral Results}

 In this section, we discuss our numerical results, including numerical tests of
 the analytical predictions given in the preceding section.

\subsection{Iterative algorithm for computing the width of NPT regions}

 In this subsection, we given an algorithm for computing NPT regions in the WBRM model.
 Its justification is given in Appendix.
 The algorithm consists of the following five steps.

\begin{figure}
\includegraphics[width=0.48\textwidth]{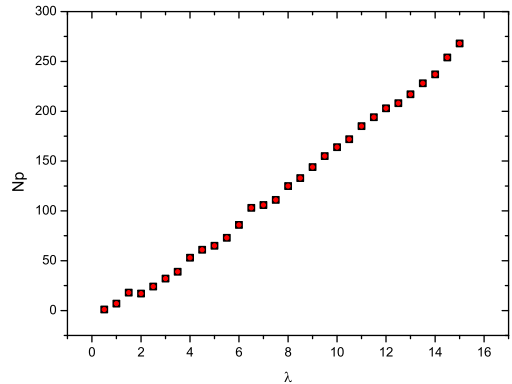}
% Here is how to import EPS art
\caption{Confirmation of the new algorithm. For each perturbation strength $\lambda$,
we randomly choose one Wigner band random matrix with $b=15$.
Black squares represent NPT width calculated with ordinary method,
and red circles represent NPT width calculated with our new iterative algorithm.
Their consistency indicates that our algorithm is correct.}
\label{confirm}
\end{figure}

Step 1: Separate the matrix $H_0$ and $V$ into blocks as
\begin{equation}\label{VpVn}
  H_0=\left(\begin{array}{cc}H_{0p}&0\\0&H_{0n}\end{array}\right),
  V=\left(\begin{array}{cc}V_p&*\\ *&V_n\end{array}\right),
\end{equation}
in which $p$ and $n$ stand for up and down. $H_{0p}$ and $V_p$ are both $[E_\alpha]\times [E_\alpha]$
square matrices, and $H_{0n}$ and $V_n$ are $N-[E_\alpha]$ dimensional square matrices.
This option ensures that $E_\alpha-H_{0p}$ is positive definite and $E_\alpha-H_{0n}$
is negative definite.

Step 2: Compute $S_p$ and $S_n$,
\begin{equation}\label{Sp}
  S_p=\frac{1}{\sqrt{E_\alpha-H_{0p}}}\lambda V_p\frac{1}{\sqrt{E_\alpha-H_{0p}}},
\end{equation}
\begin{equation}\label{Sn}
  S_n=\frac{1}{\sqrt{-(E_\alpha-H_{0n})}}\lambda V_n\frac{1}{\sqrt{-(E_\alpha-H_{0n})}}.
\end{equation}

Step 3: Compute $I+S_p$, where $I$ is the identity matrix.
 Use Guassian elimination method to eliminate elements in the lower triangle part of $I+S_p$.
 In doing this, when all the lower-triangle elements in the $i$th column are eliminated,
 a diagonal element $y_{i+1}$ is gained in the $(i+1)$-th row and column.
 We terminate the procedure when a diagonal element $y_{i_{c1}}<0$ is obtained.

\begin{figure}
\includegraphics[width=0.48\textwidth]{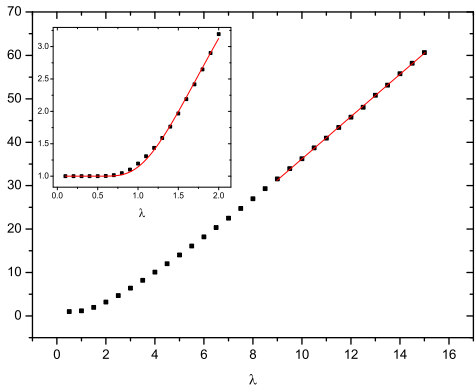}
\caption{Average NPT width towards perturbation strength $\lambda$ for Wigner-band random
matrices with band width $b=1$. When $\lambda$ is small, we fit the scatters with Eq.(\ref{small}).
When $\lambda$ is relatively large, we fit linear.}
\label{b=1NPT}
% Here is how to import EPS art
\end{figure}

Step 4: Apply the same procedure as in step 3 to $I-S_p$ and obtain a $y_{i_{c2}}<0$.
 Then, $p_1=max\{i_{c1},i_{c2}\}$.

Step 5: For $I+S_n$ and $I-S_n$, we eliminate the elements in the upper triangle
by the Gaussian-elimination method starting from the last line. Similar to steps 3 and 4,
 we take $p_2$ as the smaller $i_c$ obtained from the two matrices.
%where $i_c$ is the row and column index for the diagonal element $y_{i_c}$
%that we find less than zero at the first time in the elimination procedure.
 The NPT width is then given by $N_p=p_2-p_1$.

 The algorithm discussed above is applicable only for the case where $p_2-p_1>b$,
 in which the elements in $*$ of $V$ do not take part in the elimination.
 Practically, for a given band matrix, we first apply the above-discussed iterative algorithm
 to compute $p_2$ and $p_1$. If $N_p=p_2-p_1\le b$, then, we turn to
 the ordinary method to compute $N_p$.

\begin{figure}
\includegraphics[width=0.48\textwidth]{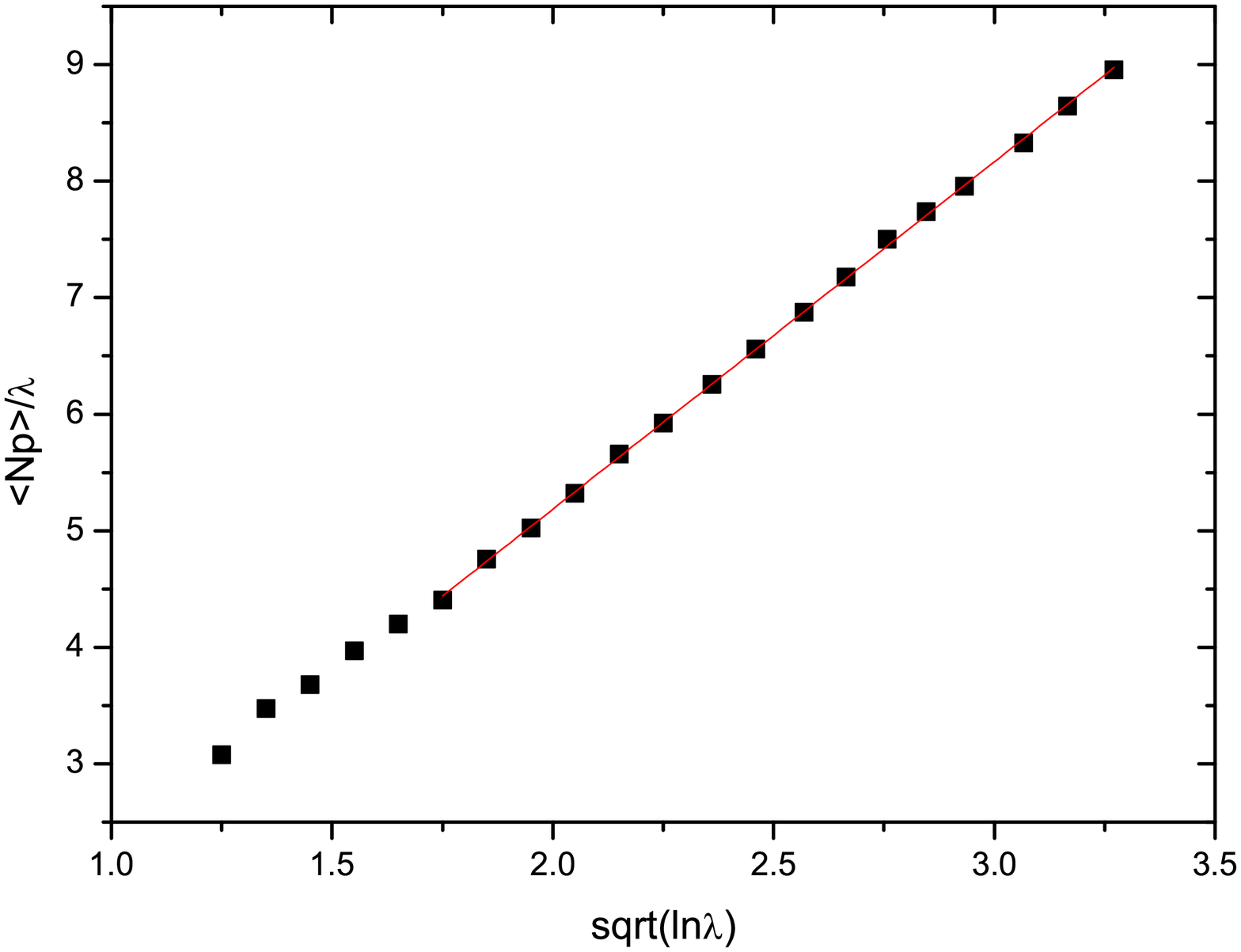}
\caption{Average NPT width towards perturbation strength $\lambda$ for Wigner band
random matrices with band width $b=1$ when $\lambda$ is very large.
We change the axis to make a linear fit with Eq.(\ref{large}).}
\label{b=1NPTLarge}
% Here is how to import EPS art
\end{figure}

 Finally, we discuss efficiency of the above-discussed algorithm.
 Similar to the method of Gaussian elimination, the algorithm has a time complexity $O(Nb^2)$.
 Furthermore, the elimination is from top and bottom to the middle of the matrix. 
 Thus, when $\lambda$ is relatively large and $p_1$ and $p_2$ are far away from middle, 
 we do not need to eliminate the total $N$ lines to find $p_1$ and $p_2$. 
 We remark that our algorithm is particularly useful for large $\lambda$.
 For small $\lambda$ and not large $b$, $N_p <b$ and the ordinary method of computing
 $N_p$ is not quite time-consuming.
 Detailed discussions of the algorithm is given in Appendix B and C.

\begin{figure}
\includegraphics[width=0.48\textwidth]{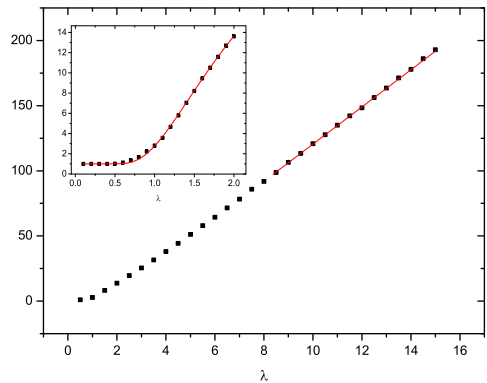}
\caption{Average NPT width towards perturbation strength $\lambda$ for
Wigner band random matrices with band width $b=8$. When $\lambda$ is small,
we fit the scatters with Eq.(\ref{small}). When $\lambda$ is relatively large, we fit linear.}
\label{b=8NPT}
% Here is how to import EPS art
\end{figure}

 We have numerically tested the validity of the above-discussed algorithm (see Fig.\ref{confirm}).

\subsection{Variation of the NPT width with perturbation strength}

 According the analytical study discussed in Sec.\ref{sect-analytical}, for $b=1$,
 we have the following picture for variation of $\la N_p\ra$, the average width of
 the NPT regions of EFs.
 That is, in the perturbation regime from weak to somewhat medium, specifically, for $\lambda \lesssim 1$,
 it follows Eq.(\ref{small}).
 One should note that, since the level spacing of the unperturbed Hamiltonian is one,
 the perturbation is not weak at $\lambda =1$.
 While, for large $\lambda$, the behavior is given by Eq.(\ref{large}).
 For $\lambda$ not very large, Eq.(\ref{large}) shows that the width increases almost linearly
 with $\lambda$, but, effect of the logarithm term should be seen for sufficiently large $\lambda$.

 As shown in Fig.\ref{b=1NPT}, Eq.(\ref{small}) works quite well for $\lambda $ below $2$.
 For $\lambda$ beyond $9$, $N_p$ has a good linear behavior in agreement with the prediction of
 Eq.(\ref{large}) for $\lambda$ not very large.
 For very large $\lambda$, the contribution of $\sqrt{\ln\lambda}$ in Eq.(\ref{large})
 becomes unnegligible, as shown in Fig.\ref{b=1NPTLarge}.

 We further increase the width $b$ of the Hamiltonian matrices.
 As shown in Fig.\ref{b=8NPT}, our predictions also works well in this case.
 For $b>1$, the parameters $C$ and $\beta$ in Eq.(\ref{small})
 and parameter $C'$ in Eq.(\ref{large}) are fitting parameters.

\begin{figure}
\includegraphics[width=0.48\textwidth]{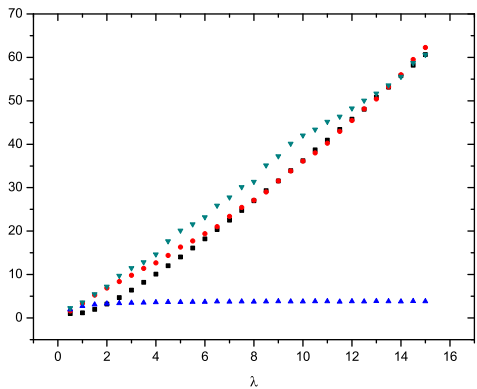}
\caption{Average NPT width(black squares), LDOS halfwidth(dark green triangles),
half width of eigenstates(red circles) and localization length(blue triangles)
for Wigner band random matrices with band width $b=1$.}
\label{b=1comp}
% Here is how to import EPS art
\end{figure}

\subsection{NPT width and half-width of LDOS}

 As discussed previously, the half-width of LDOS should be closely related to the width of NPT regions
 of EFs [cf.Eq.(\ref{wL-Np})].
 In this subsection, we numerically test this prediction.

\begin{figure}
\includegraphics[width=0.48\textwidth]{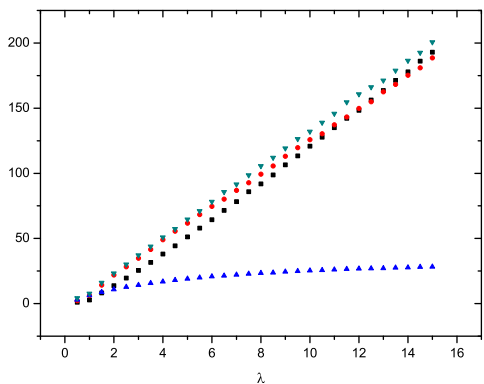}
\caption{Average NPT width(black squares), LDOS halfwidth(dark green triangles),
half width of eigenstates(red circles) and localization length(blue triangles) for Wigner band random matrices with band width $b=8$.}
\label{b=8comp}
% Here is how to import EPS art
\end{figure}

\begin{figure}
\includegraphics[width=0.48\textwidth]{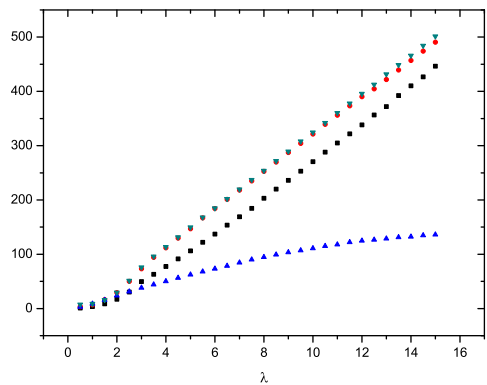}
% Here is how to import EPS art
\caption{Average NPT width(black squares), LDOS halfwidth(dark green triangles),
half width of eigenstates(red circles) and localization length(blue triangles)
for Wigner band random matrices with band width $b=50$.}
\label{b=50comp}
\end{figure}

 In Fig.\ref{b=1comp}, Fig.\ref{b=8comp}, and Fig.\ref{b=50comp},
 we give variation of $N_p$, $w_L$, and $w_{F}$ with the perturbation
 strength $\lambda$.
 We also plot the average of the localization length $L_\alpha$, denoted by $L$, where
\begin{equation}
   L_\alpha =1/\sum_{j}|C_{\alpha j}|^4.
\end{equation}
 At large $\lambda$, both $w_L$ and $w_{EF}$ are much larger than $L$, which is a phenomenon called
 localization in the energy shell.

\begin{figure}
\includegraphics[width=0.48\textwidth]{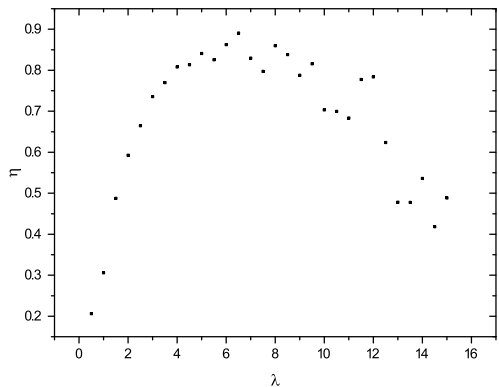}
% Here is how to import EPS art
\caption{Value of $\eta$ against $\lambda$ for $b=8$, $N=500$.
$\eta\approx 1$ for middle-strength perturbation, and is small in weak or strong perturbation.}
\label{eta}
\end{figure}

 Finally, we discuss Eq.(\ref{wL-Np}).
 We plot $\eta$ versus $\lambda$ in Fig.\ref{eta}.
 It is seen that $\eta\approx 1$ in the middle region of $\lambda$,
 and is small in the weak or strong perturbation regimes.

\section{Conclusions}

 In this paper, we have studied the NPT and PT parts of EFs in the WBRM model.
 It is shown that the width of the LDOS can be estimated, if the width of NPT regions of EFs
 are known.
 For $b=1$, we have derived explicit expressions for the average width of NPT regions,
 as functions of the perturbation strength $\lambda$, for $\lambda \gtrsim 1$ and for large $\lambda$.
 Numerically, we have found that these expressions are still valid for $b>1$.
 We have also developed a algorithm, which can efficiently compute the width of NPT regions
 at large $\lambda$, with a time complexity less than $O(b^2  N)$.

 \acknowledgements

 This work was partially supported by the Natural Science Foundation of China under Grant
 Nos.~11275179 and 11535011, the National Key Basic Research Program of China under Grant
 No.~2013CB921800, and the National Training Programs of Innovation and
 Entrepreneurship for Undergraduates.

\appendix

 \section{asymptotic property of $H(t)$ and properties of $P(n)$}
 \label{app1}

\begin{figure}
\includegraphics[width=0.48\textwidth]{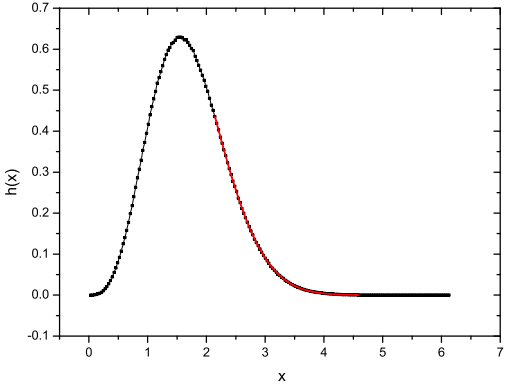}
% Here is how to import EPS art
\caption{Density function $h(x)$ of the maximum absolute value of eigenvalues of a
standard random matrix with $b=1$. Guassian fit of the large-$\lambda$ region is shown in red curve.}
\label{hx1}
\end{figure}

\begin{figure}
\includegraphics[width=0.48\textwidth]{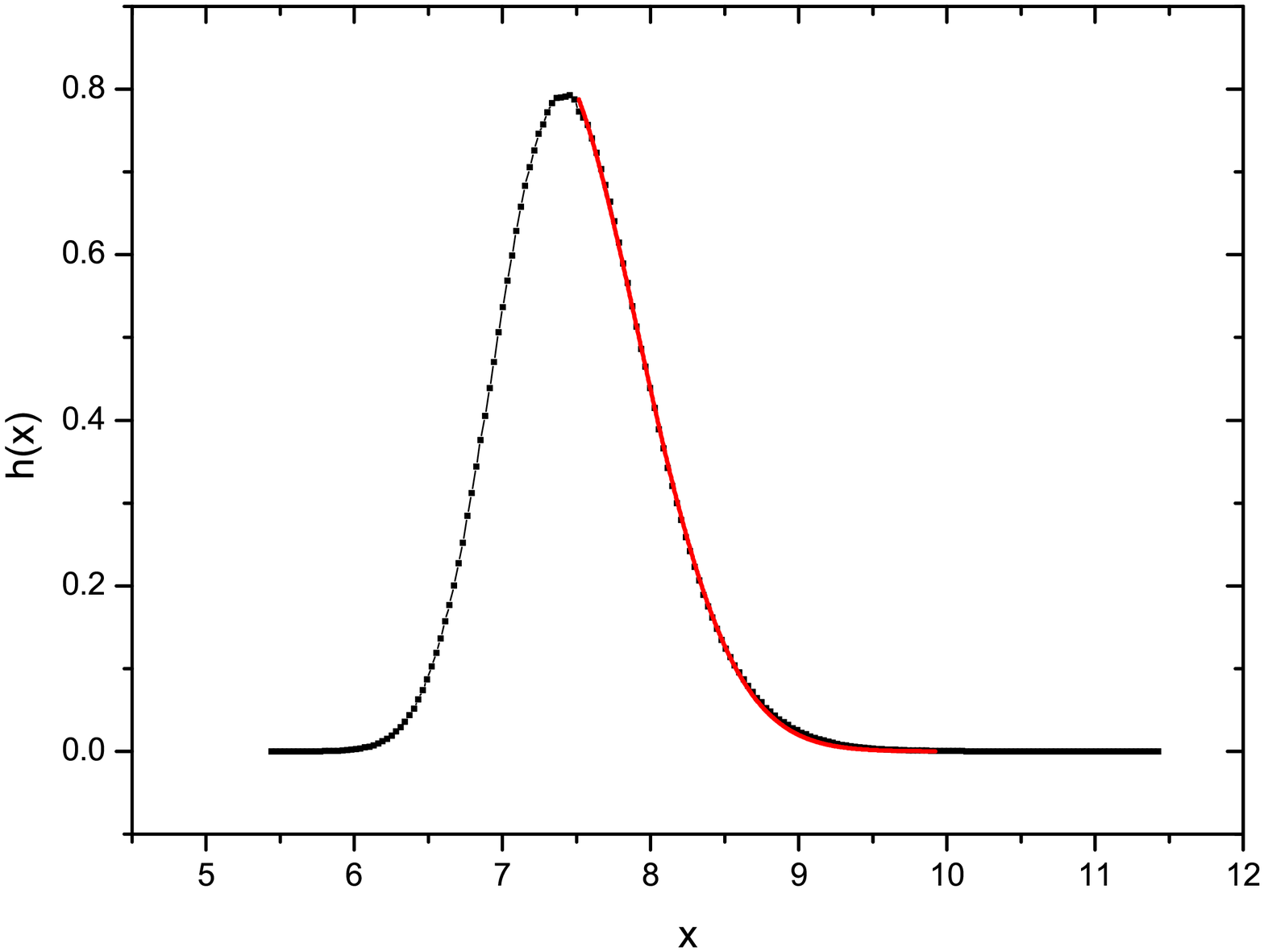}
% Here is how to import EPS art
\caption{Density function $h(x)$ of the maximum absolute value of eigenvalues of a standard random
matrix with $b=8$. Guassian fit of the large-$\lambda$ region is shown in red curve.}
\label{hx8}
\end{figure}

 Previously, we define $h(t)$ as the distribution function of the maximal
 modulus of eigenvalues of a standard $m$-dimensional random matrix $M_s$,
 whose elements take the form
\begin{equation}
  M_{s,ij}=\left\{
\begin{array}{cc}
v_{ij}\sim N(0,1), &1\le |i-j| \le b\\
0,&\rm{others}.
\end{array}
\right.
\end{equation}
and a corresponding cumulative distribution function $H(t)$.
We now explicitly show the curves of $h(t)$ when $b=1$ and $b=8$ obtained by a Monte Carlo simulation
in Fig.\ref{hx1} and Fig.\ref{hx8}. We fit the region of large $t$ by Guassian formula and
show the asymptotic property of $h(t)$ that
\begin{equation}
  h(t)\sim e^{-t^2} (t\rightarrow\infty).
\end{equation}
Making use of
\begin{equation}
  H(x)=1-\int_{x}^{+\infty}h(t)dt
\end{equation}
and
\begin{equation}\label{Gaussint}
  \int_{x}^{+\infty}t^{-n}e^{-t^2}dt\sim x^{-n-1}e^{-x^2},
\end{equation}
we obtain
\begin{equation}
  1-H(t)\sim \frac{e^{-t^2}}{t} (t\rightarrow\infty).
\end{equation}
We take first order approximation that $\ln(1-H(t))\approx-H(t)$ and use Eq.(\ref{Gaussint}) to obtain
\begin{equation}
  \int_{x}^{+\infty} \ln H(t)dt\sim \frac{e^{-x^2}}{x^2},
\end{equation}
which is just Eq.(\ref{integral}).

\section{Verification of the iterative algorithm}

To develop our algorithm, we first make a similarity transformation to $U$ to create a symmetrical matrix.
We first consider the upper part of $U$ called $U_p$ ,for $Q$ naturally split $U$ into
two independent parts. We will denote the correspondingly upper part of $V$
and $H_0$ as $V_p$ and $H_{0p}$. Note that $Q$ and $1/(E_\alpha-H_0)$ commute,
so we rearrange $U^n_p$ as
\begin{equation}
  U^n_p=Q\frac{1}{\sqrt{E_\alpha-H_{0p}}} S_p^{n-1}\frac{1}{\sqrt{E_\alpha-H_{0p}}}\lambda Q,
\end{equation}
where
\begin{equation}
  S_p=\frac{1}{\sqrt{E_\alpha-H_{0p}}}\lambda V_p\frac{1}{\sqrt{E_\alpha-H_{0p}}}.
\end{equation}
Then the condition $s(U_p)<1$ is equivalent to $s(S_p)<1$, with $S_p$ obviously symmetrical
and with the same band width.

\par Now we recall some results in linear algebra.\cite{GBW} A real symmetrical matrix
has real eigenvalues, and condition $s(S_p)<1$ is equivalent to $I\pm S_p$ are both
positive definite, where $I$ is the identity matrix.
The sufficient and necessary condition for a symmetrical matrix to be positive definite
is that its ordered main subdeterminants(denoted by $d_i$) are all positive.
We will calculate the ordered main subdeterminants of $I\pm S_p$ by elementary row transformations.
To be clear, we first deal with the case where $b=1$.

\par In this simpler case, one can prove that if $S_p$ has a positive eigenvalue $x_0$,
it must have a corresponding eigenvalue $-x_0$, so we only need to verify the condition
under which $I+S_p$ is positive definite.
Recall the elementary row transformation matrices $T_{ij}(\mu)$, with all diagonal elements $1$
and only nonvanishing element $\mu$ in the $i$th column and $t$th row, and multiplying
$T_{ij}(\mu)$ from the left gives an elementary row transformation that does not change
any its subdeterminant. Suppose the elements of $S_p$ take the form
\begin{equation}
  S_p(i,i+1)=S_p(i+1,i)=\xi_i(i=1,2,\cdots,p_1-2),
\end{equation}
then $d_1=1$, and we can continuously multiply $T_{i,i+1}(\mu_i)$ from the left to
eliminate the element of $I+S_p$ in lower triangle to obtain $d_{i+1}$.
For instance, our first step is that
\begin{widetext}
\begin{equation}
  T_{12}(-\xi_1)(I+S_p)=\left(\begin{array}{ccccc}1& & & & \\-\xi_1&1& & &
  \\ & &1& & \\ & & &\ddots& \\ & & & &1\end{array}\right)\left(\begin{array}{ccccc}1&\xi_1& & &
  \\\xi_1&1&\xi_2 & & \\ &\xi_2&1& & \\ & & &\ddots&\xi_{p_1-2} \\ & & &\xi_{p_1-2} &1
  \end{array}\right)=\left(\begin{array}{ccccc}1&\xi_1& & & \\0&1-\xi^2_1&\xi_2 & &
  \\ &\xi_2&1& & \\ & & &\ddots&\xi_{p_1-2} \\ & & &\xi_{p_1-2} &1\end{array}\right),
\end{equation}
\end{widetext}
so $\mu_1=-\xi_1$, and $d_2=1-\xi^2_1$. If $d_2>0$, then we multiply $T_{23}(-\xi_2/(1-\xi^2_1))$
from the left to obtain $d_3$. Now we introduce the sequence $\{y_i\}$ to denote the new
diagonal element before $i$th elimination, then $y_1=1,y_2=d_2/d_1=1-\xi^2_1,\cdots,y_i=d_i/d_{i-1},
\cdots$.
We can construct a recursive formula for $y_{i+1}$ by the $i$th elimination
\begin{widetext}
\begin{equation}
 T_{i,i+1}(-\frac{\xi_{i+1}}{y_i})\left(\begin{array}{cccccccc}y_1&\xi_1 & & & & & &
 \\ &y_2&\xi_2 & & & & & \\ & &\ddots&\ddots & & & & \\ & & &y_i&\xi_{i}& & &
 \\ & & &\xi_{i} &1&\xi_{i+1} & & \\ & & & &\xi_{i+1}&1 & &\\ & & & & & &\ddots &\xi_{p_1-2}
 \\ & & & & & &\xi_{p_1-2} & 1 \end{array}\right)=\left(\begin{array}{cccccccc}y_1&\xi_1 & & & & & &
 \\ &y_2&\xi_2 & & & & & \\ & &\ddots&\ddots & & & & \\ & & &y_i&\xi_{i}& & &
 \\ & & &0&y_{i+1}&\xi_{i+1} & & \\ & & & &\xi_{i+1}&1 & &\\ & & & & & &\ddots &\xi_{p_1-2}
 \\ & & & & & &\xi_{p_1-2} & 1 \end{array}\right),
\end{equation}
\end{widetext}
from which we obtain
\begin{equation}\label{recursion}
  y_{i+1}=1-\frac{\xi^2_{i}}{y_i}(i \ge 1);y_1=1.
\end{equation}
Finally we obtain our algorithm to calculate $p_1$ for the case $b=1$.
As $p_1$ is the maximal row number that ensures all ordered main subdeterminants $d_i$ of $I+S_p$
positive, we require that $\forall i \le p_1-1,y_i>0$. Given a matrix $S$, we continuously apply
Eq.(\ref{recursion}), until we obtain a $y_{i_c}<0$, then $p_1=i_c$.
Similarly, if we set $y_N=1$, we can obtain $p_2$ by adopting the same recursive formula
for $y_{i-1}$, then obtain $N_p=p_2-p_1$.
 We can also verify this algorithm by path summation in Appendix C.
 Now we expand our algorithm to the cases where $b\ge 2$.
 In this case, our way above to calculate ordered subdeterminants $d_i$ is still applicable.
 Every time we eliminate elements in a column in lower triangle of $I\pm S_p$
 by elementary row transformations $b$ times, we obtain a higher ordered subdeterminant.
 We end the iteration until we find a negative $i_c$-order subdeterminant.
 Only difference lies in that we need to both compute $i_c$ for $I\pm S_p$, and choose $p_1$
 as the smaller one.

 Similarly, we can use the algorithm calculate $p_2$, then we obtain the NPT width $N_p=p_2-p_1$.

\section{path summation derivation for the iterative algorithm}
\label{app2}

We provide a new picture for our recursive formula Eq.(\ref{recursion}).
Recall that NPT width is the minimal dimension of subspace $P$ such that the
generalized Brillouin-Wigner perturbation expansion Eq.(\ref{alpha-ovs}) converges.
Decompose $|\alpha_{s}\rangle$ as summation of NPT eigenstates,
\begin{equation}
  |\alpha_{s}\rangle=\sum_{i=p_1}^{p_2}t_i|i\ra,
\end{equation}
then convergence of Eq.(\ref{alpha-ovs}) is equivalent to convergence of
\begin{equation}\label{Cij}
  C_{ij}=\la j |i\ra+\sum_{m=1}^{n-1}\la j|T^{m-1}|i\ra+\la j|T^n|i\ra+\cdots
\end{equation}
for any $i\in [p_1,p_2]$, and any $j$. Note that the definition of $T$ (Eq.(\ref{T-alpha}))
consists of a projection operator $Q$, so we only need to check the convergence of Eq.(\ref{Cij})
for the case where $j\notin [p_1,p_2]$. Using Eq.(\ref{T-alpha}) and Eq.(\ref{Cij}),
we can write the explicit formula for $C_{ij}$,
\begin{eqnarray}\label{Cijexpansion}
  C_{ij}=\frac{\lambda V_{ji}}{E_\alpha-E_j}+\sum_{k_1\in Q} \frac{\lambda V_{jk_1}}{E_\alpha-E_j}
  \frac{\lambda V_{k_1i}}{E_\alpha-E_{k_1}} \nonumber \\
  +\sum_{k_1,k_2\in Q} \frac{\lambda V_{jk_1}}{E_\alpha-E_j}
  \frac{\lambda V_{k_1k_2}}{E_\alpha-E_{k_1}}\frac{\lambda V_{k_2i}}{E_\alpha-E_{k_2}}+\cdots.
\end{eqnarray}
By definition of NPT width, we need to find the maximal dimension of $Q$ such that Eq.(\ref{Cijexpansion})
 converges. For simplicity, we only consider the case where $b=1$. Let we call the chain
 $k_0=j\rightarrow k_1\rightarrow\cdots\rightarrow k_n\rightarrow i=k_{n+1}$ a $(n+1)$-order path,
 then the $n$th term in Eq.(\ref{Cijexpansion}) is a summation over all $n$-order paths.
 We denote $\lambda V_{{k_i}{k_{i+1}}}/(E_\alpha-E_{k_i})$ as $f(k_i\rightarrow k_{i+1})$,
 then, as $b=1$, it vanishes unless $|k_{i+1}-k_i|=1$.

 For $b=1$, $Q$ subspace is naturally split into to two separate parts,$[1,p_1-1]$ and $[p_2+1,N]$.
 Then the summation Eq.(\ref{Cijexpansion}) vanishes unless $i=p_1$ or $i=p_2$,
 where the summation only covers paths in one side of $Q$ subspace.
 Two sides of $Q$ subspaces are similar, so let us consider the case where $i=p_1$.
 We claim that convergence of Eq.(\ref{Cijexpansion}) for $j=p_1-1$ implies convergence
 for any $j\in[1,p_1-2]$. This is because if for one of $j\in[1,p_1-2]$ the path summation
 Eq.(\ref{Cijexpansion}) diverges, then for any path $j\rightarrow\cdots\rightarrow i$,
 we can construct a path $j+1\rightarrow j\rightarrow\cdots\rightarrow i$,
 and summation over all these paths also diverges because it is just $f(j+1\rightarrow j)$
 times the former summation, meaning that $C_{{p_1}{j+1}}$ also diverges. Iterate the reasoning
 above leads to contradiction to our condition that $C_{{p_1}{p_1-1}}$ converges.
 Therefore later we only consider the case where $i=p_1$ and $j=p_1-1$.

 Since the only way to $p_1$ is through $p_1-1$, We rewrite Eq.(\ref{Cijexpansion}) as
\begin{equation}
  C_{{p_1}{p_1-1}}=f(p_1-1\rightarrow p_1)A(p_1-1\rightarrow p_1-1),
\end{equation}
where $A(j\rightarrow j)$ represents the path summation over paths starting at $j$ and
ending at $j$, without passing through $j+1$. Now we classify the paths from $p_1-1$ to $p_1-1$
by the number of times the path includes $p_1-1$ in the middle.
Suppose all paths(excluded the beginning) that include $p_1-1$ one time contribute $g(p_1-1)$
to the path summation, then all paths that include $p_1-1$ $n$ times contribute $g^n(p_1-1)$. Then
\begin{eqnarray}\label{Ap}
  A(p_1-1\rightarrow p_1-1)&=&\sum_{n=1}^{\infty}g^n(p_1-1) \nonumber \\
  &=&\frac{1}{1-g(p_1-1)}-1,
\end{eqnarray}
which converges only when $|g(p_1-1)|<1$. Next we consider the paths that contribute to $g(p_1-1)$.
We may first go to $p_1-2$, then return to $p_1-1$, then the path contributes
$f(p_1-1\rightarrow p_1-2)f(p_1-2\rightarrow p_1-1)$ to the summation.
By definition, all paths that goes to $p_1-2$ then returns to $p_1-2$
contributes to a factor $A(p_1-2\rightarrow p_1-2)$, then
\begin{eqnarray}\label{gp}
 g(p_1-1)&=&f(p_1-1\rightarrow p_1-2)f(p_1-2\rightarrow p_1-1) \nonumber \\
 &\times&(1+A(p_1-2\rightarrow p_1-2)).
\end{eqnarray}
Using Eqs.(\ref{Ap}), (\ref{gp}), we obtain
\begin{eqnarray}
  g(p_1-1)&=&f(p_1-1\rightarrow p_1-2)f(p_1-2\rightarrow p_1-1) \nonumber \\
  &\times&\frac{1}{1-g(p_1-2)}.
\end{eqnarray}
The reasoning above applies to any $j\in[1,p_1-1]$. Note that $g(1)=0$ by definition,
then we obtain the recursive formula
\begin{equation}
  g(j+1)=f(j+1\rightarrow j)f(j\rightarrow j+1)\frac{1}{1-g(j)};g(1)=0.
\end{equation}
Direct calculation of matrix elements shows that
\begin{equation}
  f(j+1\rightarrow j)f(j\rightarrow j+1)=\xi^2_j,
\end{equation}
in which $\xi^2_j$ shares the same meaning with that in Eq.(40). Now let $y_i=1-g(i)$,
then Eq.(B8) becomes
\begin{equation}
  y_{i+1}=1-\frac{\xi^2_i}{y_i};y_1=1,
\end{equation}
which is exactly Eq.(\ref{recursion}).
\par In this approach, path summation converges if and only if $g(p_1-1)<1$, i.e, $y_{p_1-1}>0$.
Therefore, as we proceed our iteration, if we find a $y_{i_c}<0$, then $i_c=p_1$,
which is consistent with our previous derivation by elementary row transformation.
$p_2$ can be derived by similar approach.

% \nonindent \textbf{Supplementary Information} is linked to the online version


\begin{thebibliography}{99}
\bibitem{Lau95} B.Lauritzen, \textit{et al.}, Phys. Rev. Lett. \textbf{74}, 5190 (1995).
\bibitem{Jac97} Ph. Jacquod, D.L. Shepelyansky, O. P. Sushkov, Phys. Rev. Lett. \textbf{78}, 923 (1997)
\bibitem{Carlos98} Carlos Mejia-Monasterio, \textit{et al.}, Phys. Rev. Letts. \textbf{81}, 5189 (1998)
\bibitem{Flm} V.V.Flambaum and F. M. Izrailev, Phys. Rev. E \textbf{61}, 2539(2000); {\it ibid.},
\textbf{64}, 026124 (2001)
\bibitem{Paski12} P. R. Zangara, \textit{et al.}, Phys. Rev. A \textbf{86}, 012322 (2012).
\bibitem{Santos12}   Lea F. Santos, F.Borgonovi and F.M. Izrailev, Phys. Rev. Lett.
\textbf{108}, 094102 (2012).
\bibitem{Santos14} E. J. Torres-Herrera and Lea F. Santos, Phys. Rev. A \textbf{89}, 043620 (2014);
{\it ibid.}, \textbf{90}, 033623 (2014).
 \bibitem{GBW}  Wen-ge Wang, Phys.~Rev. E {\bf 61}, 952 (2000); {\it ibid.} {\bf 65}, 036219 (2002).

\bibitem{CCGI96} G. Casati, B.V. Chirikov, I. Guarneri, and F.M. Izrailev,
 Phys.Lett.A {\bf 223}, 430 1996.

\bibitem{pre-98} W.-G. Wang, F.M.~Izrailev, and G.~Casati, Phys. Rev. E \textbf{57}, 323 (1998).
 \bibitem{pre00} W.-g. Wang, Phys.~Rev.~E {\bf 61}, 952 (2000).

\bibitem{WBRM} E. Wigner, Ann. Math. {\bf 62,} 548 (1955); {\bf 65,} 203 (1957).

 \end{thebibliography}
 \end{document}